\documentclass[a4paper,fleqn,usenatbib,useAMS]{mnras}

\usepackage{mathptmx}
\usepackage{soul}
\usepackage{cancel}
\usepackage[T1]{fontenc}
\usepackage{ae,aecompl}
\usepackage{multirow}

\usepackage{graphicx}	
\usepackage{times,amsmath,amssymb,afterpage,float,morefloats}
\usepackage{hyperref}
\providecommand{\e}[1]{\ensuremath{\times 10^{#1}}}
\newcommand{\kms}{\,km\,s$^{-1}$}

\title[Periodic class II methanol masers in G339.986-0.425]{Discovery of periodic class II methanol masers associated with G339.986-0.425 region}
\author[J. P. Maswanganye et al.]
  {J.P. ~Maswanganye$^{1,2}$\thanks{E-mail: jabulani@hartrao.ac.za},
     D.J. ~van der Walt$^2$, S. ~Goedhart$^{2,3}$,  and M.J. ~Gaylard$^{1}$\thanks{Deceased 2014 August 14.} \\
$^1$Hartebeesthoek Radio Astronomy Observatory, PO Box 443, Krugersdorp, 1740, South Africa\\
$^2$ Center for Space Research,  North-West University, Potchefstroom campus, Private Bag X6001, Potchefstroom, 2520, South Africa\\
$^3$ SKA SA, 3rd Floor, The Park, Park Rd, Pinelands, 7405, South Africa
}
\date{Accepted  2015 December 17;      Received  2015 December 08;      in original form 2015 July 16}

\pagerange{\pageref{firstpage}--\pageref{lastpage}}
\pubyear{2015}
\begin{document}


\pagerange{\pageref{firstpage}--\pageref{lastpage}} \pubyear{2015}

\maketitle

\label{firstpage}

\begin{abstract}

Ten new class II methanol masers from the 6.7-GHz Methanol Multibeam survey catalogues III and IV were selected for a monitoring programme at both 6.7 and 12.2 GHz with the 26m Hartebeesthoek Radio Astronomy Observatory (HartRAO) radio telescope for two years and nine months, from August 2012 to May 2015. In the sample, only masers associated with G339.986-0.425 were found to show periodic variability at both 6.7 and 12.2 GHz. The existence of periodic variation was tested with four independent methods. The analytical method gave the best estimation of the period, which was 246 $\pm$ 1 days. The time series of G339.986-0.425 show strong correlations across velocity channels and between the 6.7 and 12.2 GHz masers. The time delay was also measured across channels and shows structure across the spectrum which is continuous between different maser components.

\end{abstract}

%

\section{Introduction}

The two brightest class II methanol maser transitions (6.7- and 12.2-GHz) are now known to be uniquely associated with a very early phase of high-mass star formation and to be reliable tracers of high-mass young stellar objects (YSOs) \citep{ellingsen2006}. It has been postulated from the maser morphologies that they could be tracing a circumstellar disc or torus \citep[e.g., ][]{norris1993,phillips1998,minier2000,bartkiewicz2011}, or collimated outflows associated with the high-mass YSO  \citep[e.g., ][]{minier2000}. Multiple epoch and monitoring observations of the 6.7- and 12.2-GHz masers in many star forming regions have shown that the masers are variable and in some cases highly variable \citep[e.g., ][]{caswell1995a,caswell1995b,macleod1996}. Since the masers are sensitive to the physical conditions in the circumstellar environment, the observed variability of the masers must tell us something about changes either in the physical conditions where the masers operate or in the conditions of the region where the seed photons, which stimulate the emission, originate. Maser monitoring can therefore be a useful probe of the physical conditions of those parts of the circumstellar environments that affect the masers.

In a monitoring programme to test the nature of the variability in 54 methanol sources, seven sources showed periodic variations \citep{goedhart2003,goedhart2004,goedhart2009,goedhart2013}. The discovery of periodic variability in the 6.7- and  12.2-GHz methanol masers reported by \citet{goedhart2003,goedhart2004,goedhart2009,goedhart2013} was unexpected as it had not been previously reported in any maser transitions associated with high-mass star formation regions. Further monitoring programmes have been conducted independently and confirmed the existence of periodic variations in eight more methanol masers \citep{araya2010,szymczak2011,szymczak2015,fujiswa2014,maswanganye2015}, bringing the total number of periodic masers to 15.
 
In an attempt to explain the origin of the periodic masers, five mechanisms have been proposed, three of which suggest that the variations are due to changes in the temperature of the dust grains that are responsible for the infrared radiation field that pumps the masers. These are (i) the rotation of spiral shocks in the gaps of discs around young binary stars \citep{sobolev2007,parfenov2014}, (ii) periodic accretion in a circumbinary system \citep{araya2010} and (iii) the pulsation of a YSO
\citep{inayoshi2013,sanna2015}. The remaining two mechanisms propose that the origin of variations is the result of changes in the flux of seed photons (or radio continuum) owing to either (iv) a colliding wind binary (CWB) system \citep{vanderwalt2009,vanderwalt2011} or (v) a very young low mass companion blocking the
UV radiation from the high-mass star in an eclipsing binary \citep{maswanganye2015}. There is currently no direct observational evidence to support any of the five proposed mechanisms.

The observed light-curves from the fifteen periodic methanol masers show a wide variety of shapes. \citet{maswanganye2015} classified the light-curves in four groups and argued that the light-curves could be a trace of the origin of the observed periodic masers. It remains to be seen if there are more groups or whether these are the only ones. It was also proposed by \citet{maswanganye2015} that there could be a total of 34 $\pm$ 10 periodic sources in the 6.7-GHz Methanol Multibeam (MMB) survey catalogues I, II, III, and IV \citep{caswell2010,caswell2011,green2010,green2012}. This implies that monitoring more sources in the  6.7-GHz MMB survey catalogues, including the recently published catalogue V \citep{breen2015}, could result in the discovery of more periodic sources. Such new discoveries could result in a better characterisation of the observed light-curves, and their possible relation to the period if such a relation does exist at all. There are also possibilities for new mechanisms to explain the origin of periodicity because some of the existing proposals use the light-curves as the guide for their arguments \citep[e.g., ][]{vanderwalt2011,vanderwalt2009,araya2010,maswanganye2015}.

In order to increase our sample of periodic methanol masers and improve the understanding of the methanol masers' periodic variations and characterise light-curves, more sources should be monitored. Ten sources from the 6.7-GHz MMB survey catalogues III and IV were considered for monitoring at both 6.7- and  12.2-GHz. From the sample of ten, only the masers associated with G339.986-0.425 show periodic variations at both 6.7- and  12.2-GHz. This paper describes the details of the methods used for data collection, reduction, calibration, results and analysis.

%
\section{Observations and data reduction}

%
\subsection{Source selection and HartRAO monitoring}

From the 6.7-GHz MMB survey catalogues III and IV, ten sources associated with methanol masers were selected for the monitoring programme with the 26m HartRAO radio telescope at both 6.7- and  12.2-GHz. The list of sources in the sample is given in Table \ref{tab:sources_list}. The selection criterion for these sources was the same as given by \citet{maswanganye2015}.

\begin{table*}
 \centering
 \begin{minipage}{180mm}
    \caption{Monitored sources associated with class II methanol masers from the 6.7-GHz MMB survey catalogues III and IV. Columns two and three are Right Ascension (RA) and Declination (Dec), respectively, reported by either \citet{caswell2010} or \citet{green2010}. Column four gives the central frequencies. Columns five and six give the velocity range within which masers were found. The flux densities reported from two epochs in the 6.7-GHz MMB survey are given in columns seven, from MX mode or beam-switching technique data, and in column eight, from a survey cube (SC) data \citep{green2009}. The total time-span is given in column nine. Column ten gives a number of epochs the source was observed, excluding observations which were considered as bad data. In column eleven, is the catalogue number in which the 6.7-GHz maser was selected from.}
  \label{tab:sources_list}
  \begin{tabular}{@{}lcccccccccc@{}}
  \hline
   Source Name  &  \multicolumn{2}{c}{Equatorial Coordinates}   & Frequency   &   \multicolumn{2}{c}{Velocity range} & \multicolumn{2}{c}{MMB survey flux}  & Monitoring Window  & Number & MMB \\
   ( l,   b )   &  RA (2000)   & Dec. (2000)                    &    &  $V_L$   & $V_H$       & MX data & SC data & Start - End  & of &  Catalogue   \\
   ($^{\circ}$, $^{\circ}$) & (h  m  s) & ($^{\circ}$ $^{\prime}$   $^{\prime \prime}$) & (GHz) & \multicolumn{2}{c}{ (\kms)} & (Jy) & (Jy) & (MJD) & Epochs & Number\\  
   \hline
   
\multirow{2}{*}{G312.071+0.082} & \multirow{2}{*}{14 08 58.20}  & \multirow{2}{*}{-61 24 23.8} & 6.7  & -38.0 & -28.0 & 67.86    &  83.2  & 56160 - 56710  & 35 & \multirow{2}{*}{IV}   \\
                                &                               &                              & 12.2 & -38.0 & -28.0 & 67.86    &  83.2  & 56179 - 56339  & 6 &    \\
\multirow{2}{*}{G320.780+0.248} & \multirow{2}{*}{15 11 23.48}  & \multirow{2}{*}{-57 41 25.1} & 6.7  & -11.0 & -3.0 &  34.92    &  24.44 & 56250 - 56801  & 39 & \multirow{2}{*}{IV}   \\
                                &                               &                              & 12.2 & -8.0  & -3.0 &  34.92    &  24.44 & 56151 - 56339  & 13 &  \\
\multirow{2}{*}{G329.719+1.164} & \multirow{2}{*}{15 58 07.09}  & \multirow{2}{*}{-51 43 32.6} & 6.7  & -85.0 & -70.0 & 7.69     &  24.44 & 56151 - 57115  & 66 & \multirow{2}{*}{IV}   \\
                                &                               &                              & 12.2 & -     & -     &  -       &  -     & 56171 - 56179  & 2 &   \\ 
\multirow{2}{*}{G335.426-0.240} & \multirow{2}{*}{16 30 05.58}  & \multirow{2}{*}{-48 48 44.8} & 6.7  & -54.0 & -39.0 & 66.00    &  91.20 & 56151 - 56549  & 31 &  \multirow{2}{*}{III}  \\
                                &                               &                              & 12.2 & -54.0 & -48.0 & 66.00    &  91.20 & 56160 - 56339  & 10 &   \\
\multirow{2}{*}{G337.052-0.226} & \multirow{2}{*}{16 36 40.17}  & \multirow{2}{*}{-47 36 38.4} & 6.7  & -88.0 & -73.0 & 14.00    &  23.43 & 56151 - 57104  & 61 & \multirow{2}{*}{III}  \\
                                &                               &                              & 12.2 & -     & -     &      -   &   -    & 56151 - 56160 & 2 &    \\
\multirow{2}{*}{G337.153-0.395} & \multirow{2}{*}{16 37 48.86}  & \multirow{2}{*}{-47 38 56.5} & 6.7  & -52.0 & -46.0 & 17.50    &  22.06 & 56160 - 56652 & 21 &  \multirow{2}{*}{III}  \\
                                &                               &                              & 12.2 & -     &  -    &     -    &   -    & 56151 - 56179  & 2 &  \\
\multirow{2}{*}{G337.388-0.210} & \multirow{2}{*}{16 37 56.01}  & \multirow{2}{*}{-47 21 01.2} & 6.7  & -68.0 & -50.0 & 23.00    &  24.38 & 56160 - 56800  & 35 & \multirow{2}{*}{III}  \\
                                &                               &                              & 12.2 & -     & -     &      -   &   -    & 56151 - 56151  & 1 &  \\
\multirow{2}{*}{G338.925+0.634} & \multirow{2}{*}{16 40 13.56}  & \multirow{2}{*}{-45 38 33.2} & 6.7  & -70.0 & -52.0 & 64.00    &  74.84 & 56150 - 57093  & 44 & \multirow{2}{*}{III}  \\
                                &                               &                              & 12.2 & -70.0 & -52.0 & 64.00    &  74.84 & 56151 - 56323  & 4 &  \\
\multirow{2}{*}{G339.986-0.425} & \multirow{2}{*}{16 48 46.31}  & \multirow{2}{*}{-45 31 51.3} & 6.7  & -92.0 & -86.0 & 90.00    &  69.07 & 56160 - 57175  & 66 & \multirow{2}{*}{III}  \\
                                &                               &                              & 12.2 & -92.0 & -86.0 & 90.00    &  69.07 & 56160 - 57175  & 70 &  \\
\multirow{2}{*}{G343.354-0.067} & \multirow{2}{*}{16 59 04.23}  & \multirow{2}{*}{-42 41 35.0} & 6.7  & -129.0& -114.0 & 18.00   &  20.31 & 56150 - 56710  & 23 & \multirow{2}{*}{III}  \\
                                &                               &                              & 12.2 & -     & -     &    -     &  -     & 56151 - 56160  & 2 & \\
\hline
\end{tabular}
\end{minipage}
\end{table*}

The data for the 6.7- and  12.2-GHz masers were captured using the frequency switch technique into the 1024-channel spectrometer for the 26m HartRAO radio telescope. The spectral resolutions were 0.044 and 0.048\kms\ for the 6.7- and  12.2-GHz maser observations, respectively. For any  source with the brightest peak greater than 20 Jy, a pointing observation to correct for pointing error, was made prior to the long on-source scan. The pointing observation had five scans which were at north, east, west, and south of the half power beam, and on-source. All sources were observed at least once every one to three weeks. If a source showed any form of variability, the cadence was increased to at least once every week \citep{maswanganye2015}. The typical on-source RMS noise for the 12.2-GHz masers for which there was no detection with the 26m HartRAO radio telescope was $\sim$ $1$ Jy, with the integration time between 6 and 8 minutes. In the case where a source had been observed more than once in a day, the observations were averaged into one observation. 

After pointing correction (if applicable) and data reduction, the spectra were calibrated using the point source sensitivity (PSS) derived from drift scans of Hydra A and 3C123. Observations of Hydra A and 3C123 at 6.7- and  12.2-GHz were made daily depending on the availability of the telescope and they were independent of the maser observations. In each drift scan observation, three scans: north and south of the half power beamwidth, and on-source were made. After baseline corrections to the scans, the amplitudes of scans were determined and used to calculate for the pointing corrections and PSS. The flux density of the calibrators were adopted from \citet{ott1994}. The PSS values were averaged over the period where there were no step changes. The averaged PSS minimises extrinsic variations when the spectra are scaled to Jansky.

%
\subsection{ATCA interferometry}

Interferometric data on G339.986-0.425  were obtained from the archive of the follow up observations of the 6.7-GHz MMB survey with the ATCA. The Australian National Telescope Facility (ATNF) project code is C1462 \citep{fuller2006}. Six snapshot scans were made with the 6B configuration of the 22m ATCA antennas. One scan was 2.67 minutes long and the remaining five scans were 2.58 minutes. PKS B1934-638 was only observed at the beginning of the observations for 2.91 minutes and it was used as the bandpass calibrator and flux density calibrator. For gain calibration, PKS B1646-50 was used. The data were reduced and calibrated with MIRIAD.

%

\section{Results}

All the methanol masers associated with the sources given in Table \ref{tab:sources_list} were initially observed at least four times with the 26m HartRAO radio telescope at both 6.7- and  12.2-GHz. The search for the 12.2-GHz methanol maser counterparts was conducted independently of the results of the 12.2-GHz MMB follow up catalogue II \citep{breen2012}. Of all sources which passed the test phase at either or both 6.7- and  12.2-GHz, only one source, G339.986-0.425, has shown strong variations over the monitoring window in both transitions. This section gives the results of the single dish monitoring programme for the two brightest class II methanol masers associated G339.986-0.425 and the rest of the sources which have shown weak variations are given in the appendix.

\begin{figure}
\centering
\resizebox{\hsize}{!}{\includegraphics[clip]{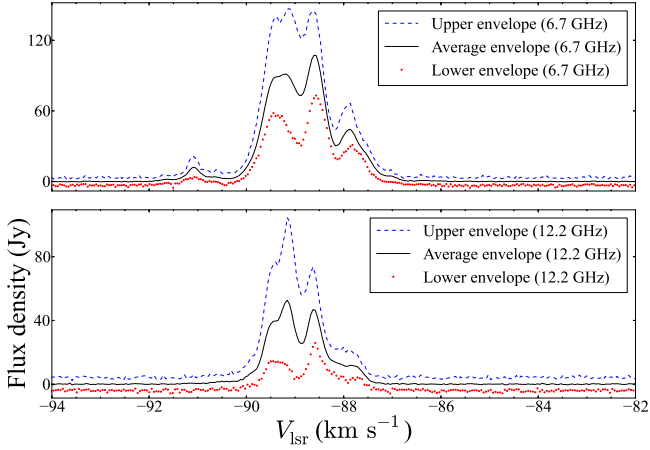}}
\caption{Spectra for the methanol masers associated with G339.986-0.425 at both 6.7- and  12.2-GHz. The velocity width is $\sim$ 5.5 and 3.5\kms\ for the 6.7- and  12.2-GHz masers, respectively.}
\label{fig:G339.986-0.425_6_and_12ghz_spectra}
\end{figure}

The spectra obtained with the 26m HartRAO radio telescope for G339.986-0.425 are shown in Figure \ref{fig:G339.986-0.425_6_and_12ghz_spectra}. For both transitions, the upper (lower) envelope was defined as the absolute maximum (minimum) flux density attained over the monitoring window measured in each channel. The average envelope was obtained by averaging the flux densities across the monitoring window in each channel. The 6.7-GHz spectra obtained with the 26m HartRAO radio telescope are similar to that reported by \citet{caswell2011}. The 6.7- and  12.2-GHz spectra are similar but there is a maser feature around -91\kms\ in the 6.7-GHz emission which is absent in the in the 12.2-GHz spectra. The 6.7-GHz masing features occurred between -91.8 and -86.5\kms, whereas 12.2-GHz maser features were between -90.5 and -87.0\kms.

\begin{figure}
\centering
\resizebox{\hsize}{!}{\includegraphics[clip]{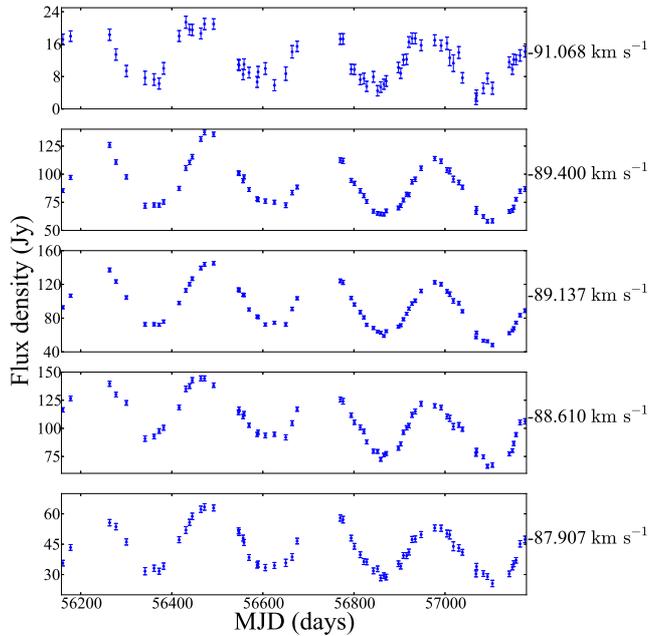}}
\caption{Time series for the methanol masers associated with G339.986-0.425 at 6.7-GHz.}
\label{fig:G339.986-0.425_67ghz_timeseries}
\end{figure}

\begin{figure}
\centering
\resizebox{\hsize}{!}{\includegraphics[clip]{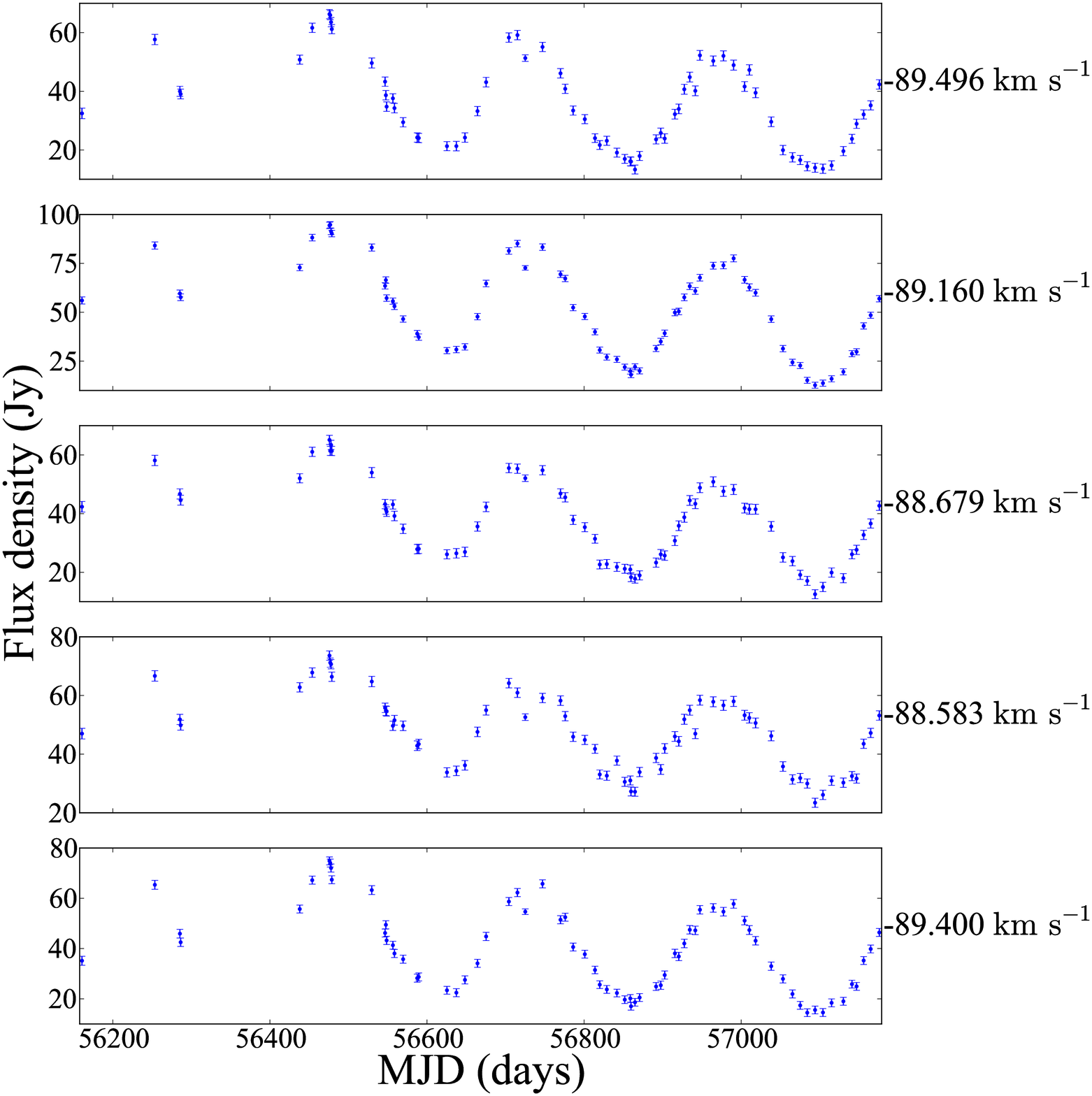}}
\caption{Time series for the methanol masers associated with G339.986-0.425 at 12.2-GHz.}
\label{fig:G339.986-0.425_12ghz_timeseries}
\end{figure}

\begin{figure}
\centering
\resizebox{\hsize}{!}{\includegraphics[clip]{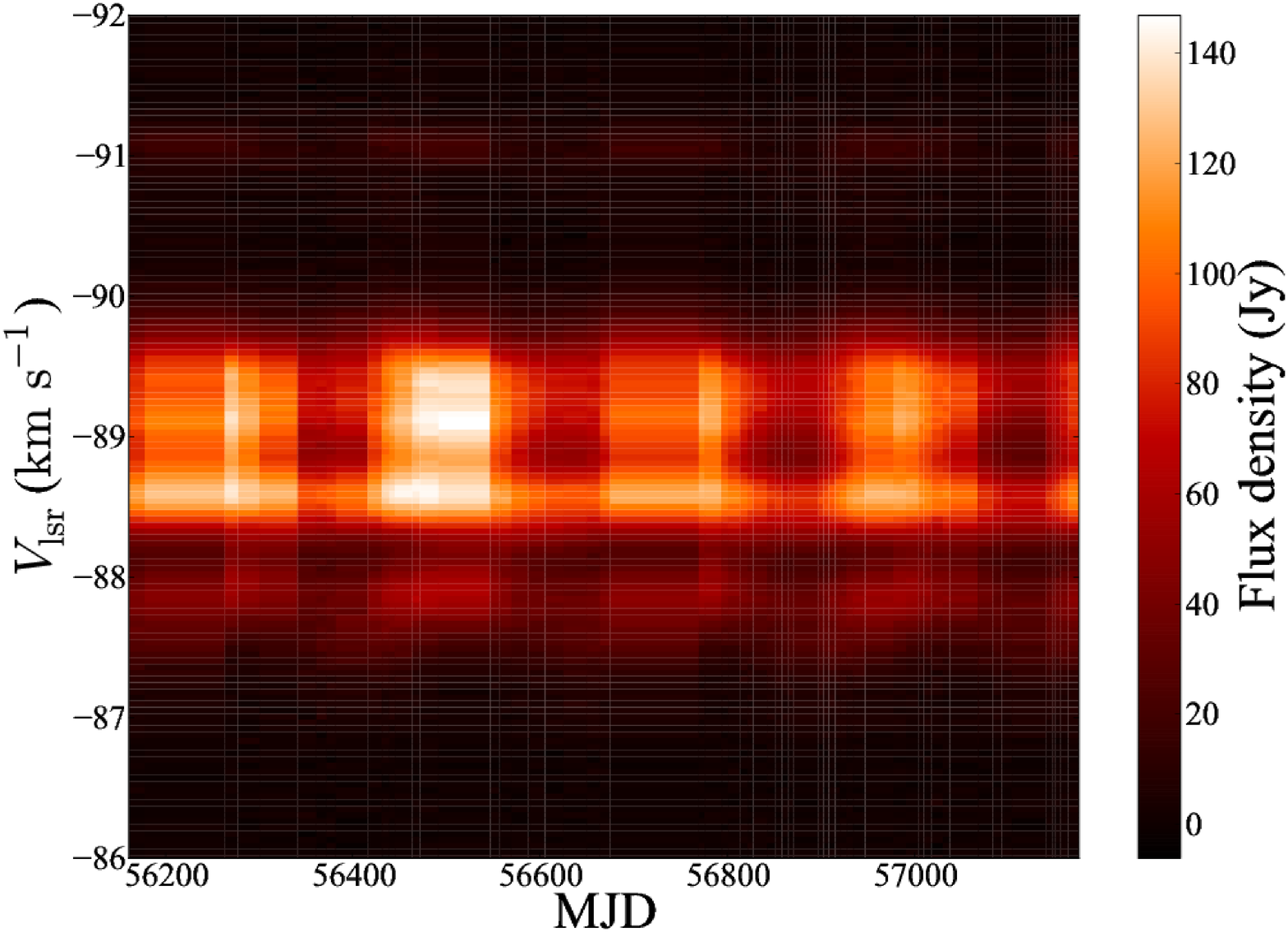}}
\caption{Time series for class II methanol masers associated with G339.986-0.425 at 6.7-GHz in 2-d colour-map form.}
\label{fig:G339.986-0.425_67_2d_timeseries}
\end{figure}

\begin{figure}
\centering
\resizebox{\hsize}{!}{\includegraphics[clip]{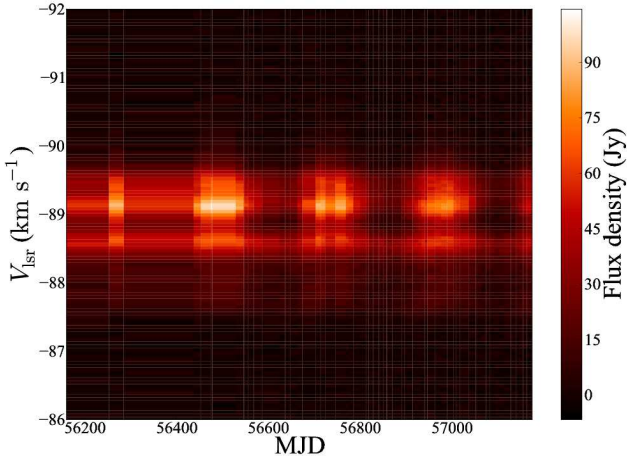}}
\caption{Time series for class II methanol masers associated with G339.986-0.425 at 12.2-GHz in 2-d colour-map form.}
\label{fig:G339.986-0.425_122_2d_timeseries}
\end{figure}

The time series for the flux density of the 6.7- and  12.2-GHz masers at the peak velocities are shown in Figures \ref{fig:G339.986-0.425_67ghz_timeseries} and \ref{fig:G339.986-0.425_12ghz_timeseries} respectively. By visual inspection, the time series show periodic variations and strong correlation between the 6.7- and  12.2-GHz masers, and within each maser peak. \citet{caswell2011} reported variability in flux densities over two epochs, but the nature of this variation was not determined. The 2-d plots in Figures \ref{fig:G339.986-0.425_67_2d_timeseries} and \ref{fig:G339.986-0.425_122_2d_timeseries} show the complete flux density variations between -92.0 and -86.0\kms\ of the 6.7- and  12.2-GHz masers, respectively. All masers within the region show similar patterns of variability, which implies that the time series at the peak velocities are representative of the general behaviour of the adjacent velocities.


%
\section{Data analysis techniques}

We searched for periodicity in the time series using the Lomb-Scargle \citep{lomb1976,scargle1982,press1989}, Jurkevich \citep{jurkevich1971} and epoch-folding using Linear-Statistics (or L-statistics) \citep{davies1990} methods. The correlations and time delays between the time series of different channels were determined using the z-transformed discrete correlation function (ZDCF) \citep{alexander1997}.

The Lomb-Scargle method is a modified classical periodogram or spectral analysis. The significance of the fundamental peak or a peak in the periodogram was tested with the false alarm probability method \citep{scargle1982}.

The epoch-folding and Jurkevich methods use a trial period to fold the time series. The epoch-folding using Linear-Statistics (or L-statistics) and Jurkevich statistics are calculated from the folded time series. The period in the epoch-folding is the location of the fundamental peak. In the Jurkevich method, the period is derived from the location of the absolute minimum.

The ZDCF method fixes one time series and moves the other across the reference time series to search for a correlation and time delay.
%
\section{Analysis}

%
\subsection{Period search}

\begin{figure}
\centering
\resizebox{\hsize}{!}{\includegraphics[clip]{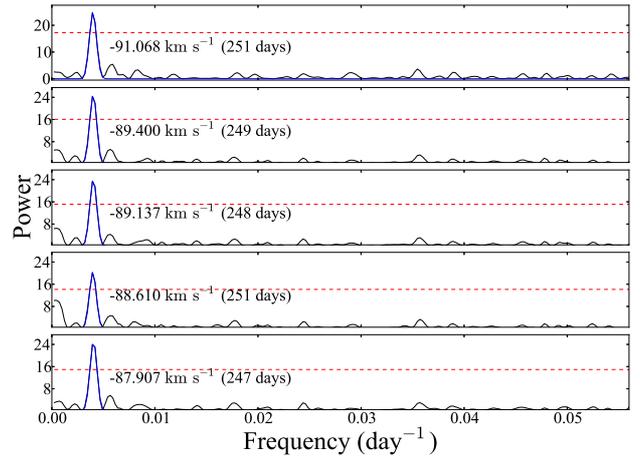}}
\caption{The Lomb-Scargle periodogram for G339.986-0.425 at 6.7-GHz. The dashed line is the cut-off power flux for false detected peak. So peaks below the cut-off power are considered to be a false alarm or false detection of the peak.}
\label{fig:G339.986-0.425_67_lomb_scargle}
\end{figure}

\begin{figure}
\centering
\resizebox{\hsize}{!}{\includegraphics[clip]{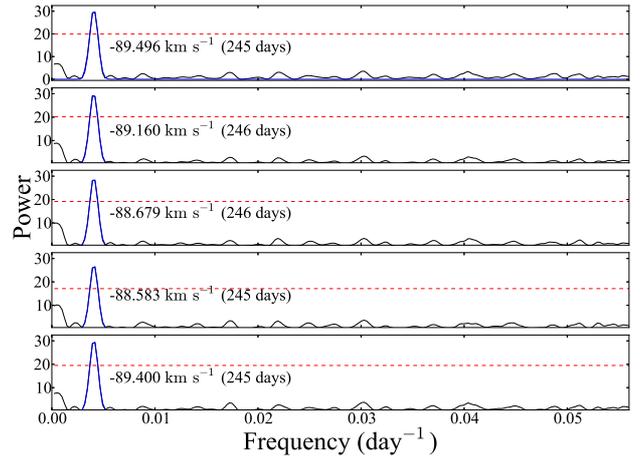}}
\caption{The Lomb-Scargle periodogram for G339.986-0.425 at 12.2-GHz.}
\label{fig:G339.986-0.425_122_lomb_scargle}
\end{figure}

\begin{figure}
\centering
\resizebox{\hsize}{!}{\includegraphics[clip]{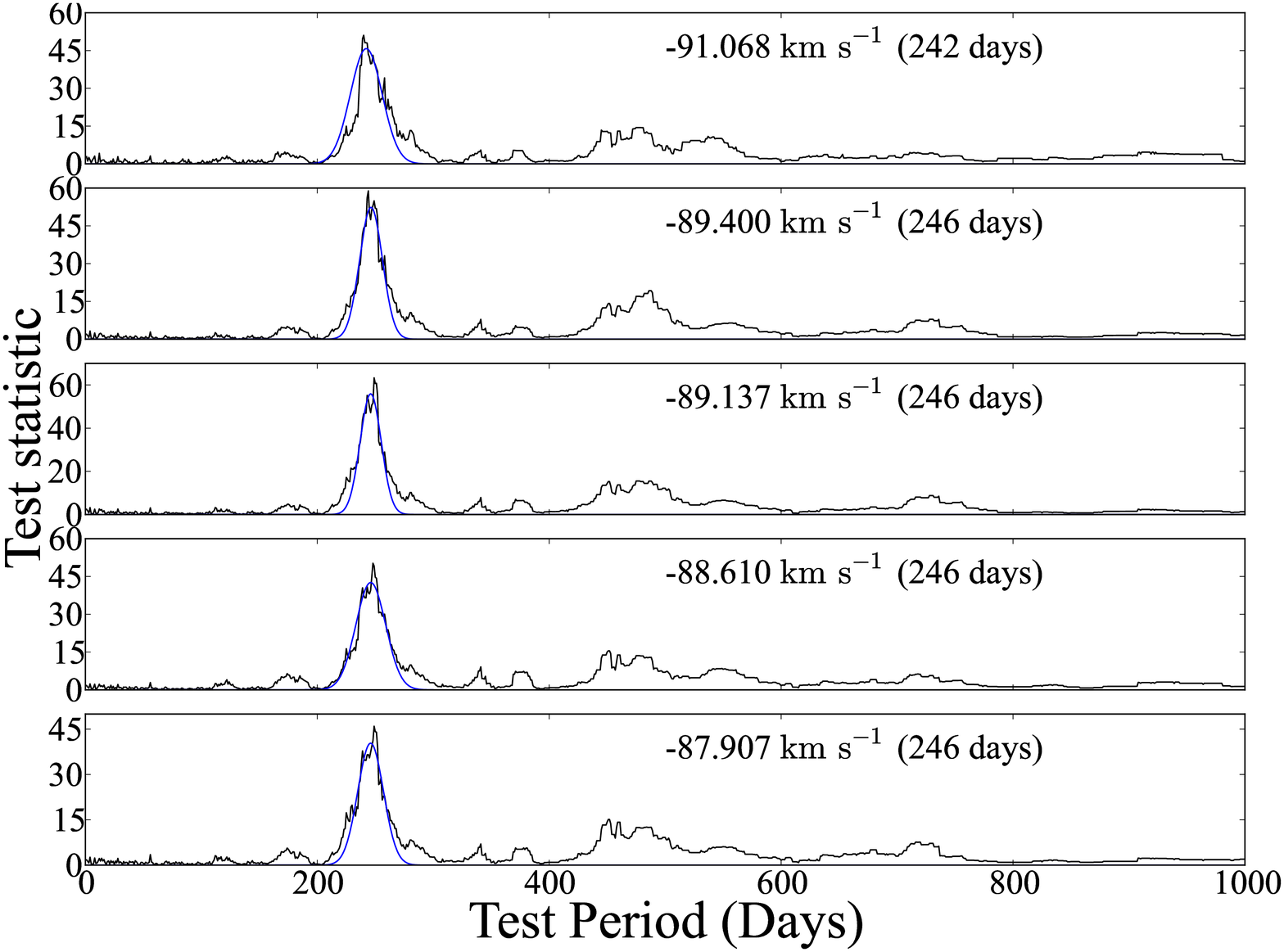}}
\caption{The epoch-folding result using L-statistics for G339.986-0.425 at 6.7-GHz. The solid (blue) line around the fundamental peak is for a fitted Gaussian function around the peak. The period and its uncertainty were extracted from the fitted Gaussian parameters.}
\label{fig:G339.986-0.425_67_epochfolding}
\end{figure}

\begin{figure}
\centering
\resizebox{\hsize}{!}{\includegraphics[clip]{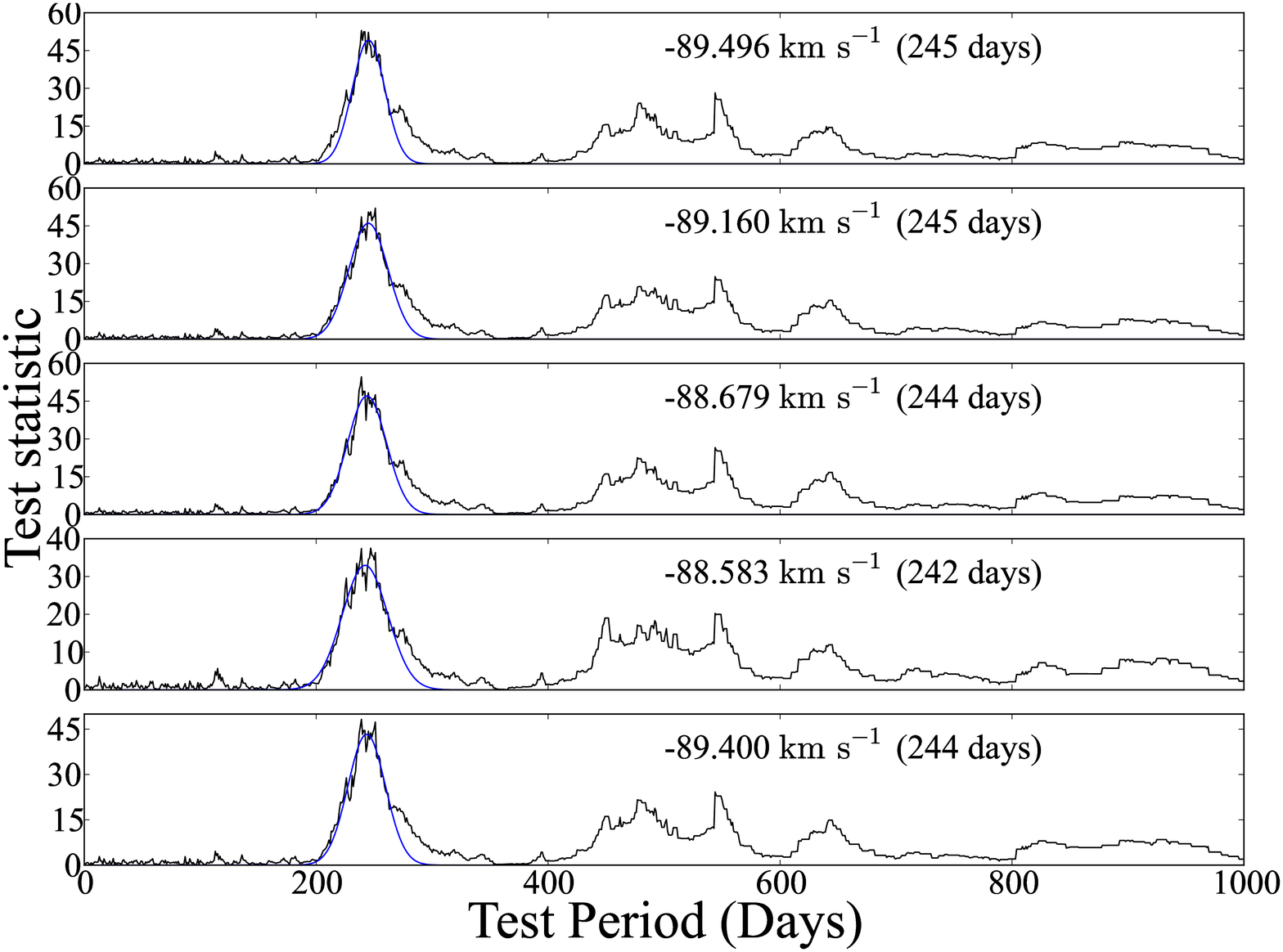}}
\caption{The epoch-folding result using L-statistics for G339.986-0.425 at 12.2-GHz.}
\label{fig:G339.986-0.425_122_epochfolding}
\end{figure}

\begin{figure}
\centering
\resizebox{\hsize}{!}{\includegraphics[clip]{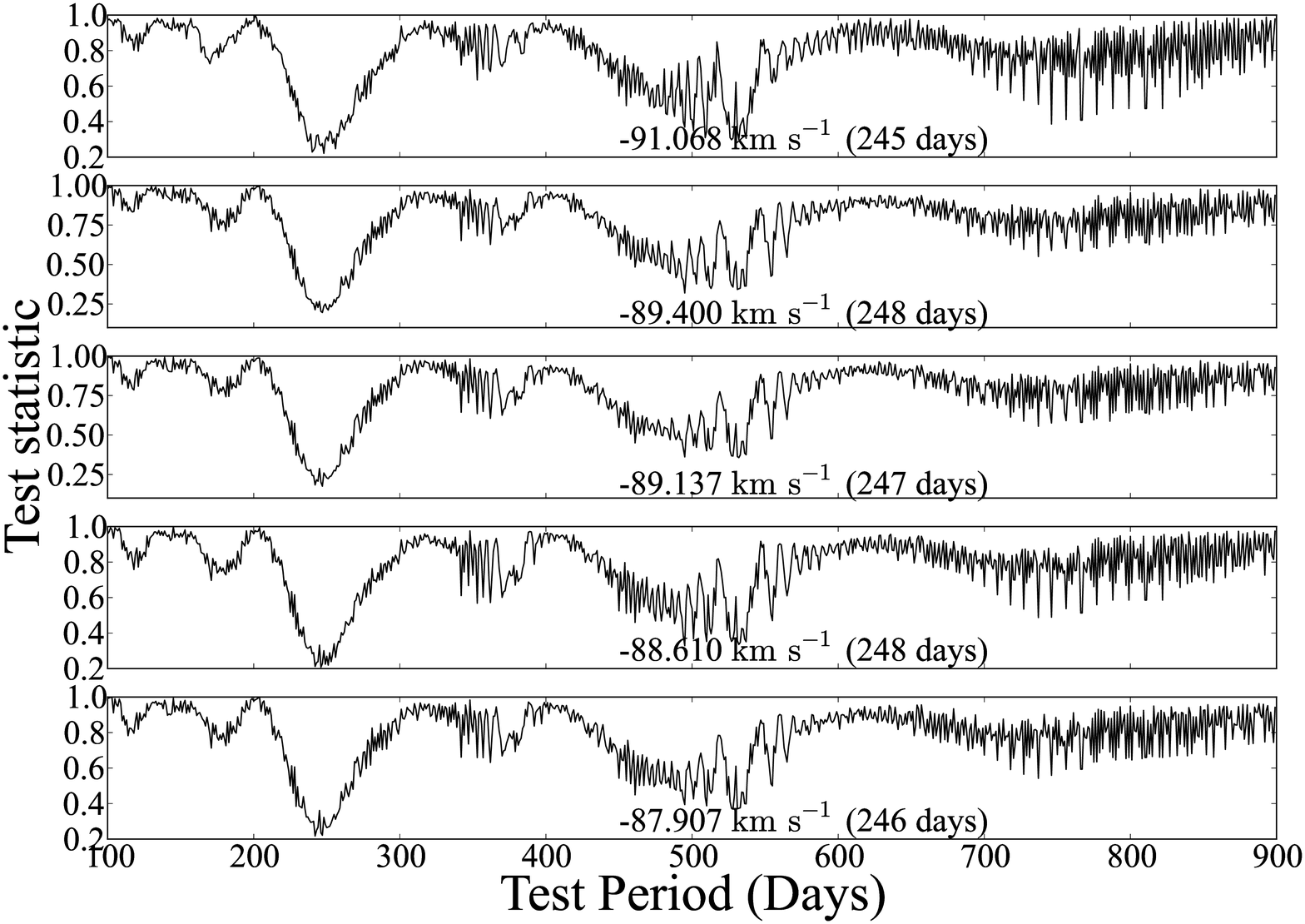}}
\caption{The Jurkevich-statistics result for G339.986-0.425 at 6.7-GHz.}
\label{fig:G339.986-0.425_67_jurkevich}
\end{figure}

\begin{figure}
\centering
\resizebox{\hsize}{!}{\includegraphics[clip]{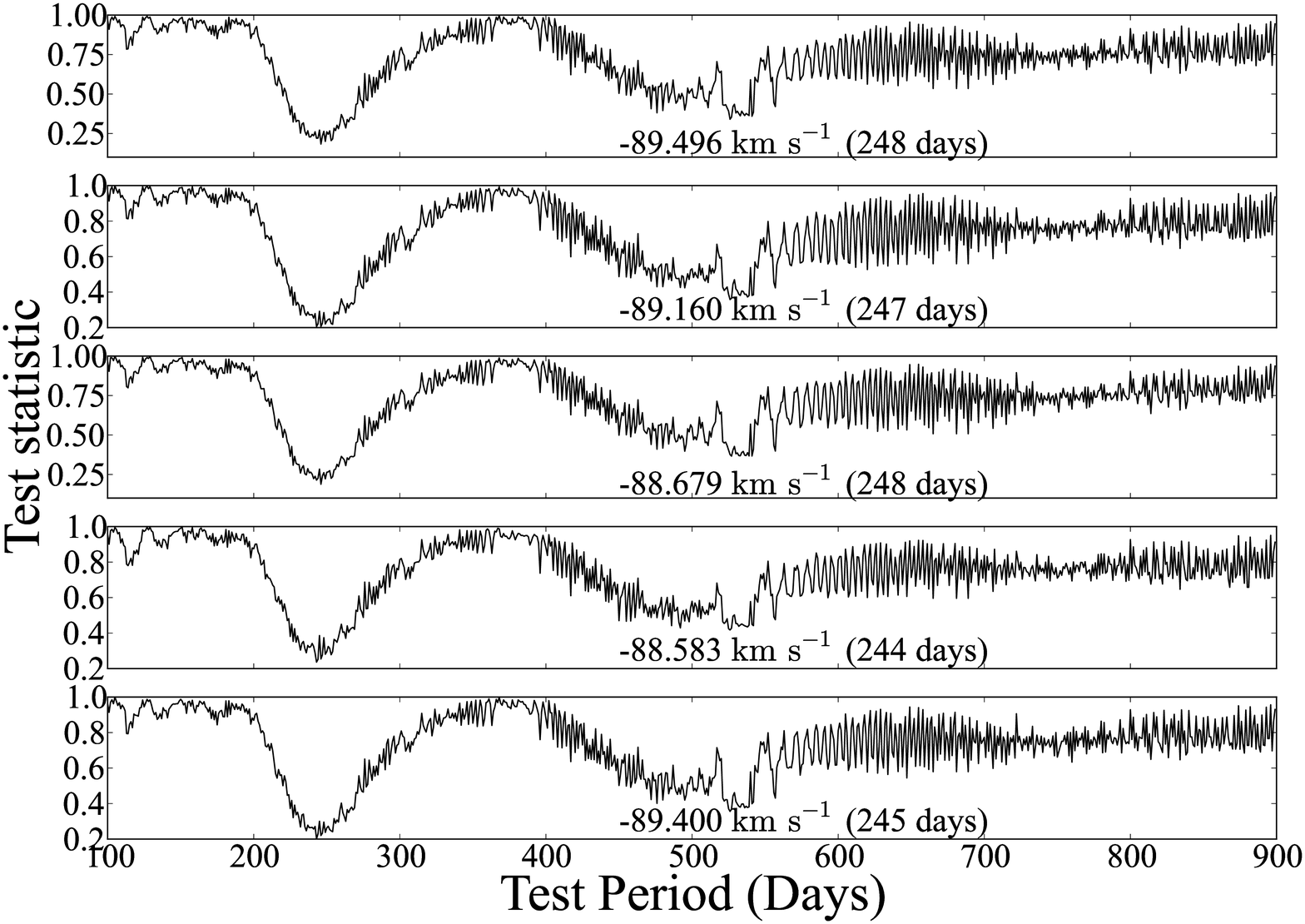}}
\caption{The Jurkevich-statistics result for G339.986-0.425 at 12.2-GHz.}
\label{fig:G339.986-0.425_122_jurkevich}
\end{figure}

Of all the sources listed in Table \ref{tab:sources_list}, only G339.986-0.425 has shown periodic variations. The Lomb-Scargle (Figures \ref{fig:G339.986-0.425_67_lomb_scargle} and \ref{fig:G339.986-0.425_122_lomb_scargle}), epoch-folding using L-statistic (Figures \ref{fig:G339.986-0.425_67_epochfolding} and \ref{fig:G339.986-0.425_122_epochfolding}) and Jurkevich (Figures \ref{fig:G339.986-0.425_67_jurkevich} and \ref{fig:G339.986-0.425_122_jurkevich}) results confirmed the existence of periodicity over the monitoring window and they also agreed in the determined period. A summary of the determined periods and their uncertainties for the peak velocities are given in Table \ref{tab:periods}. The uncertainty in the period was estimated as the half width at half maximum (HWHM) of the fundamental peak for the Lomb-Scargle and epoch-folding methods, and for Jurkevich method, it was the HWHM of the absolute minimum dip. The epoch-folding method had smaller uncertainties when compared to the other two methods.

All maser sources which were removed early in the monitoring programme with small number of observation epochs, due to low signal-to-noise-ratio with the 26m HartRAO radio telescope, were not tested for the periodicity.

\begin{table*}
 \begin{minipage}{140mm}
  \caption{Summary of the determined periods using the Lomb-Scargle, epoch-folding and Jurkevich period search methods.}
  \label{tab:periods}
  \begin{center}
  \begin{tabular}{@{}ccccc@{}}
  \hline
  Maser feature velocity & Frequency & Lomb-Scargle & Epoch-folding & Jurkevich  \\
    (\kms)        & (GHz)     & (days)       &    (days)    &   (days)   \\
    \hline   
    -91.068              &  6.7       & 250 $\pm$ 25 &  242 $\pm$ 16 & 245 $\pm$ 43 \\
    -89.400              &  6.7       & 249 $\pm$ 25 &  246 $\pm$ 12 & 248 $\pm$ 30 \\     
    -89.137              &  6.7       & 248 $\pm$ 25 &  246 $\pm$ 11 & 246 $\pm$ 29 \\
    -88.610              &  6.7       & 251 $\pm$ 25 &  246 $\pm$ 15 & 248 $\pm$ 28 \\
    -87.907              &  6.7       & 247 $\pm$ 25 &  246 $\pm$ 13 & 246 $\pm$ 26 \\
    -89.400              &  12.2      & 245 $\pm$ 29 &  245 $\pm$ 17 & 248 $\pm$ 40 \\
    -89.496              &  12.2      & 245 $\pm$ 29 &  245 $\pm$ 20 & 247 $\pm$ 45 \\
    -89.160              &  12.2      & 245 $\pm$ 29 &  244 $\pm$ 20 & 248 $\pm$ 39 \\
    -88.679              &  12.2      & 245 $\pm$ 28 &  242 $\pm$ 23 & 244 $\pm$ 43 \\
    -88.583              &  12.2      & 245 $\pm$ 29 &  244 $\pm$ 19 & 245 $\pm$ 48 \\
    \hline
\end{tabular}
\end{center}
\end{minipage}
\end{table*}

The period and uncertainty estimate from the Lomb-Scargle, epoch-folding and Jurkevich methods improve as the number of cycles increase. These three methods do not use the errors in the flux densities when calculating the periodogram or test statistic. If an analytical function which can model the time series for G339.986-0.425 can be found, then a weighted chi-square fit can be used to find a better estimate of the period and its uncertainty.

\begin{figure}
\centering
\resizebox{\hsize}{!}{\includegraphics[clip]{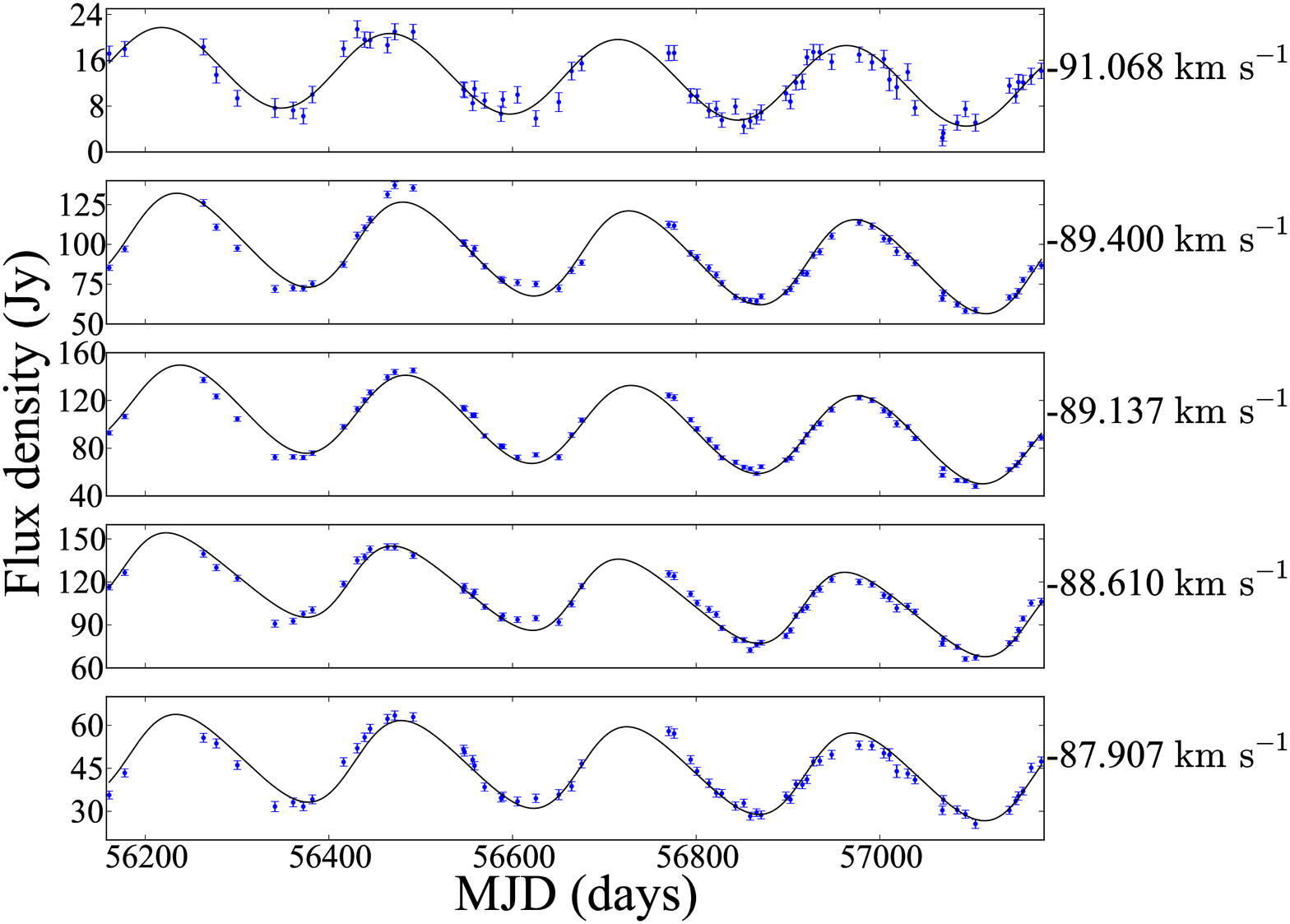}}
\caption{Time series for the methanol masers associated with G339.986-0.425 at 6.7-GHz with the best of the asymmetric cosine function (solid line).}
\label{fig:G339.986-0.425_67ghz_timeseries_fit_asymmetric_cosine}
\end{figure}

\begin{figure}
\centering
\resizebox{\hsize}{!}{\includegraphics[clip]{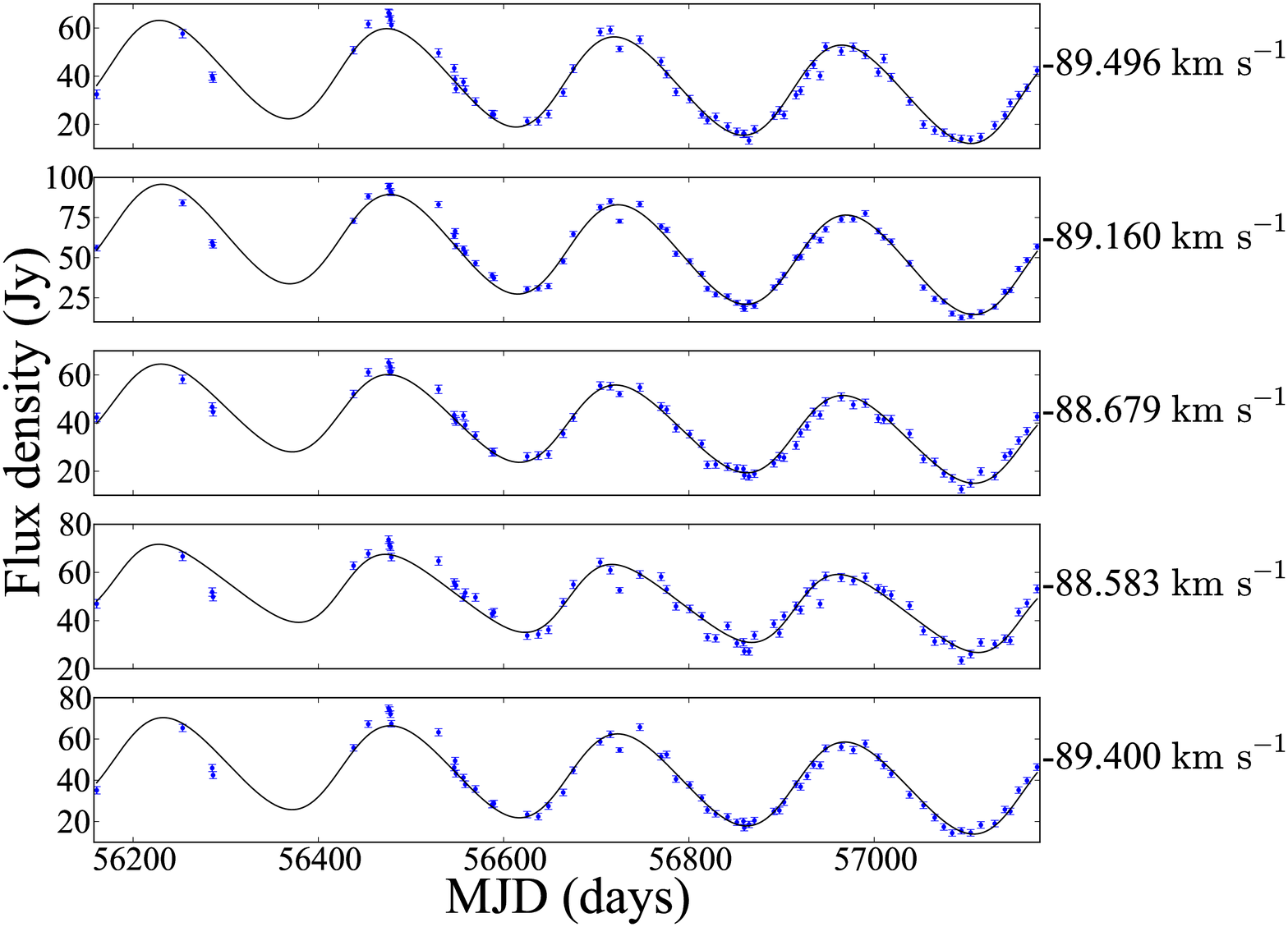}}
\caption{Time series for the methanol masers associated with G339.986-0.425 at 12.2-GHz with the best of the asymmetric cosine function (solid line).}
\label{fig:G339.986-0.425_12ghz_timeseriesfit_asymmetric_cosine}
\end{figure}

Inspection of the time series in Figures \ref{fig:G339.986-0.425_67ghz_timeseries}
and \ref{fig:G339.986-0.425_12ghz_timeseries} suggests that the light-curves are quasi-sinusoidal. The light-curves show a general behaviour of a fast rise to the local maximum which is followed by a slow decay to the local minimum. \citet{david1996} used an asymmetric cosine function to model the hydroxyl masers light-curves as seen in OH/IR stars. These light-curves are similar to what is observed in the time series of the 6.7- and  12.2-GHz masers in G339.986-0.425. The light-curves of G339.986-0.425 can be modelled as the sum of an asymmetric cosine (periodic variations) and a first order polynomial (long-term variations), which can be expressed using the following formula as
\begin{equation}
 s(t) = \frac{b\cos{\left(\omega t + \phi \right)}}{1 - f\sin{\left( \omega t + \phi \right)}} + mt + c,
 \label{eq:asymmetric_cosine}
\end{equation}
where $b$, $\omega$ ($\omega=\frac{2\pi}{P}$, $P$ is  the period), $\phi$, $f$, $m$ and $c$ are the amplitude, angular frequency, phase, eccentricity (related to the asymmetry factor $f_o$), gradient and y-intercept, respectively. The eccentricity $f$, is defined as $f=\sin{\left(\pi\left[f_o - 0.5\right]\right)}$. The asymmetry factor is given by the ratio of the rise time from the minimum to the maximum and the period. The asymmetric cosine function in equation \ref{eq:asymmetric_cosine} is symmetric if $f_o = 0.5$ which implies that the eccentricity, $f$, will be zero. The free parameters in equation \ref{eq:asymmetric_cosine} can be represented as a vector, $\mathbf{P}$, which is given as $\langle b, \omega, \phi,f, m, c \rangle$. The weighted chi-squared, $\chi^2 (t,\mathbf{P})$, can be used to determine the free parameters of the model which best fit the time series in Figures \ref{fig:G339.986-0.425_67ghz_timeseries} and \ref{fig:G339.986-0.425_12ghz_timeseries}. The non-linear weighted $\chi^2 (t,\mathbf{P})$ can be solved with the Levenberg-Marquardt algorithm \citep{levenberg1944,marquardt1963}. The results from the minimised non-linear weighted $\chi^2 (t,\mathbf{P})$ of the selected time series (Figures \ref{fig:G339.986-0.425_67ghz_timeseries} and \ref{fig:G339.986-0.425_12ghz_timeseries}) are shown in Figures \ref{fig:G339.986-0.425_67ghz_timeseries_fit_asymmetric_cosine} and \ref{fig:G339.986-0.425_12ghz_timeseriesfit_asymmetric_cosine}. Each time series in Figures \ref{fig:G339.986-0.425_67ghz_timeseries_fit_asymmetric_cosine} and \ref{fig:G339.986-0.425_12ghz_timeseriesfit_asymmetric_cosine} was fitted independently with the same initial guess period.

\begin{table*}
 \begin{minipage}{150mm}
  \caption{Summary of the determined periods, eccentricities, amplitudes and gradients from the best fitted asymmetric cosine into the selected time series shown in Figures \ref{fig:G339.986-0.425_67ghz_timeseries} and \ref{fig:G339.986-0.425_12ghz_timeseries}. In column 7, is an asymmetry factor $f_o$ which is defined as $f_o = 0.5 +\frac{\arcsin{\left( f\right)}}{\pi}$. In column 8, is the rise time from the local minimum to the local maximum which is defined as the product of the period (column 3) and $f_o$ (column 7). The uncertainties in columns 7 and 8 were derived from the error propagation formula \citep{weiss2012}.}
  \label{tab:asymetric_cosine_periods}
  \begin{center}
  \begin{tabular}{@{}cccccccc@{}}
  \hline
  Maser feature velocity & Frequency & Period           &  $f$              &     Amplitude    & Gradient                &  $f_o$             & Rise time \\
(\kms)  & (GHz)     & (days)           &                   &       (Jy)       &   (Jy/day)              &                    &    (days)   \\
    \hline   
    -91.068              &  6.7       & 248.6 $\pm$ 1.2 &  -0.05 $\pm$ 0.07 & 6.8   $\pm$ 0.3  &  -0.0042   $\pm$  0.0007 & 0.48  $\pm$ 0.02  & 120 $\pm$ 6 \\   
    -89.400              &  6.7       & 246.2 $\pm$ 0.4 &  -0.22 $\pm$ 0.02 & 27.2  $\pm$ 0.4  &  -0.023    $\pm$  0.001  & 0.429 $\pm$ 0.008 & 105 $\pm$ 2 \\     
    -89.137              &  6.7       & 245.6 $\pm$ 0.3 &  -0.15 $\pm$ 0.02 & 34.2  $\pm$ 0.4  &  -0.0347   $\pm$  0.0009 & 0.453 $\pm$ 0.007 & 111 $\pm$ 2 \\
    -88.610              &  6.7       & 246.4 $\pm$ 0.4 &  -0.33 $\pm$ 0.03 & 25.2  $\pm$ 0.4  &  -0.037    $\pm$  0.001  & 0.394 $\pm$ 0.009 & 97  $\pm$ 2 \\
    -87.907              &  6.7       & 245.5 $\pm$ 0.5 &  -0.25 $\pm$ 0.04 & 14.2  $\pm$ 0.3  &  -0.0088   $\pm$  0.0008 & 0.42  $\pm$ 0.01  & 103 $\pm$ 3 \\   
    -89.400              &  12.2      & 245.5 $\pm$ 0.5 &  -0.19 $\pm$ 0.03 & 19.1  $\pm$ 0.3  &  -0.0139   $\pm$  0.0009 & 0.440 $\pm$ 0.009 & 108 $\pm$ 2 \\
    -89.496              &  12.2      & 246.2 $\pm$ 0.3 &  -0.15 $\pm$ 0.02 & 28.9  $\pm$ 0.3  &  -0.0260   $\pm$  0.0009 & 0.451 $\pm$ 0.006 & 111 $\pm$ 2 \\
    -89.160              &  12.2      & 245.4 $\pm$ 0.6 &  -0.20 $\pm$ 0.03 & 16.6  $\pm$ 0.3  &  -0.0177   $\pm$  0.0009 & 0.43  $\pm$ 0.01  & 106 $\pm$ 3 \\
    -88.679              &  12.2      & 244.5 $\pm$ 0.6 &  -0.33 $\pm$ 0.04 & 14.2  $\pm$ 0.3  &  -0.017    $\pm$  0.001  & 0.39  $\pm$ 0.01  & 96 $\pm$ 3 \\
    -88.583              &  12.2      & 245.2 $\pm$ 0.5 &  -0.18 $\pm$ 0.03 & 20.8  $\pm$ 0.3  &  -0.0161   $\pm$  0.0009 & 0.441 $\pm$ 0.009 & 108 $\pm$ 2 \\
    \hline
\end{tabular}
\end{center}
\end{minipage}
\end{table*}

The summary of the periods, $f$, amplitudes and gradients are given in Table \ref{tab:asymetric_cosine_periods}. The errors in the free parameters were estimated from the square-root of the diagonal entries in the inverse of the covariant matrix $V_{\bf P}$ of the best fitted parameters of the model, $\hat{\sigma}_{\mathbf{P}}=\sqrt{\text{diagonal}\left( {V_{\mathbf P}}^{-1}\right)}$ \citep{press1992}. The propagation of error formula \citep{weiss2012} was used to derive the uncertainties of the periods from the angular frequencies $\omega$. The periods in Table \ref{tab:asymetric_cosine_periods} agree very well for all components. The gradients from the fits suggest a long-term linear decay and the $f$ values also confirm that there is asymmetry in the light-curves. The values of $f_o$ for G339.986-0.425 are close to that found by \citet{david1996} for the log of flux, $\log(S_{\nu})$, of the OH masers light-curves associated with OH/IR stars. The rise times of all components are similar.

A comparison of the errors obtained with the Lomb-Scargle, epoch-folding and Jurkevich methods with that obtained from the fit of the asymmetric cosine shows that for the latter the error is about a day or less while for the first three methods the errors are between 11 and 48 days. This large difference in the errors need to be clarified. It should be noted that the Lomb-Scargle, epoch-folding and Jurkevich methods do not specify a particular shape of the light-curve, which implies that the period is estimated non-parametrically, whereas fitting an asymmetric cosine function is parametric approach which will have some bias. Also, the uncertainties of the periods given in Table \ref{tab:asymetric_cosine_periods} does not necessarily depend on the number of cycles but rather number of data points. Although in general, more cycles can also imply more data points. It needs to be noticed that the error on the period from fitting the asymmetric cosine is due only to the fitting procedure. Furthermore, the fit is applied to only one instance of the sampling of the true light-curve with the implication that the error on the period does not have any statistical meaning in terms of how the light-curve was sampled. To get a more realistic estimate of the error due to statistical effects, three possible cases are considered: First, the observing times are fixed and it is argued that variations in the amplitudes at the different times, as reflected by the errors on the measured amplitudes, can possibly give rise to larger errors on the estimated period. Second, as indicated in Table \ref{tab:sources_list}, the number of samples, $n$, in the time series are respectively 66 and 70 for the 6.7- and 12.2-GHz masers. This fixed set of sampling times represent only one possible sampling of the light-curve on which an asymmetric cosine can be fitted. The question is what would we expect for the error on the period if we had more than one possible sampling of the light-curves but with the same number of time-stamps. The third possible way to estimate the error is to reliably model the observed light-curve and through a Monte-Carlo method generate synthetic light-curves on which an asymmetric cosine can be fitted. In the next couple of paragraphs we will consider each of these methods and show that a consistent estimate for the period can be obtained.

In the first investigation, a Monte-Carlo simulation was used to generate $10^{4}$ light-curves from the time series of the -89.137\kms\ maser feature. The light-curves were sampled from a Gaussian distribution with the observed flux density as the mean and the experimental error in the flux density as a standard deviation. This was done for each time-stamp in the original time series for all $10^{4}$ simulated light-curves. The periods in the simulated time series were estimated from both the weighted and unweighted fit of an asymmetric cosine. The mean and the standard deviation of the determined periods was 245.6 $\pm$ 0.3 days for the weighted fit and 246.0 $\pm$ 0.3 days for the unweighted fit. The estimated uncertainties agree with values given in Table \ref{tab:asymetric_cosine_periods}. We note, however, that the experimental errors for the time series in the -89.137\kms\ maser feature are very small relative to the measured flux densities, with a mean percentage error of $\sim$ 2 \%. This is the main reason why the calculated uncertainty in the period in this particular case is still very small.

\begin{figure}
\centering
\resizebox{\hsize}{!}{\includegraphics[clip]{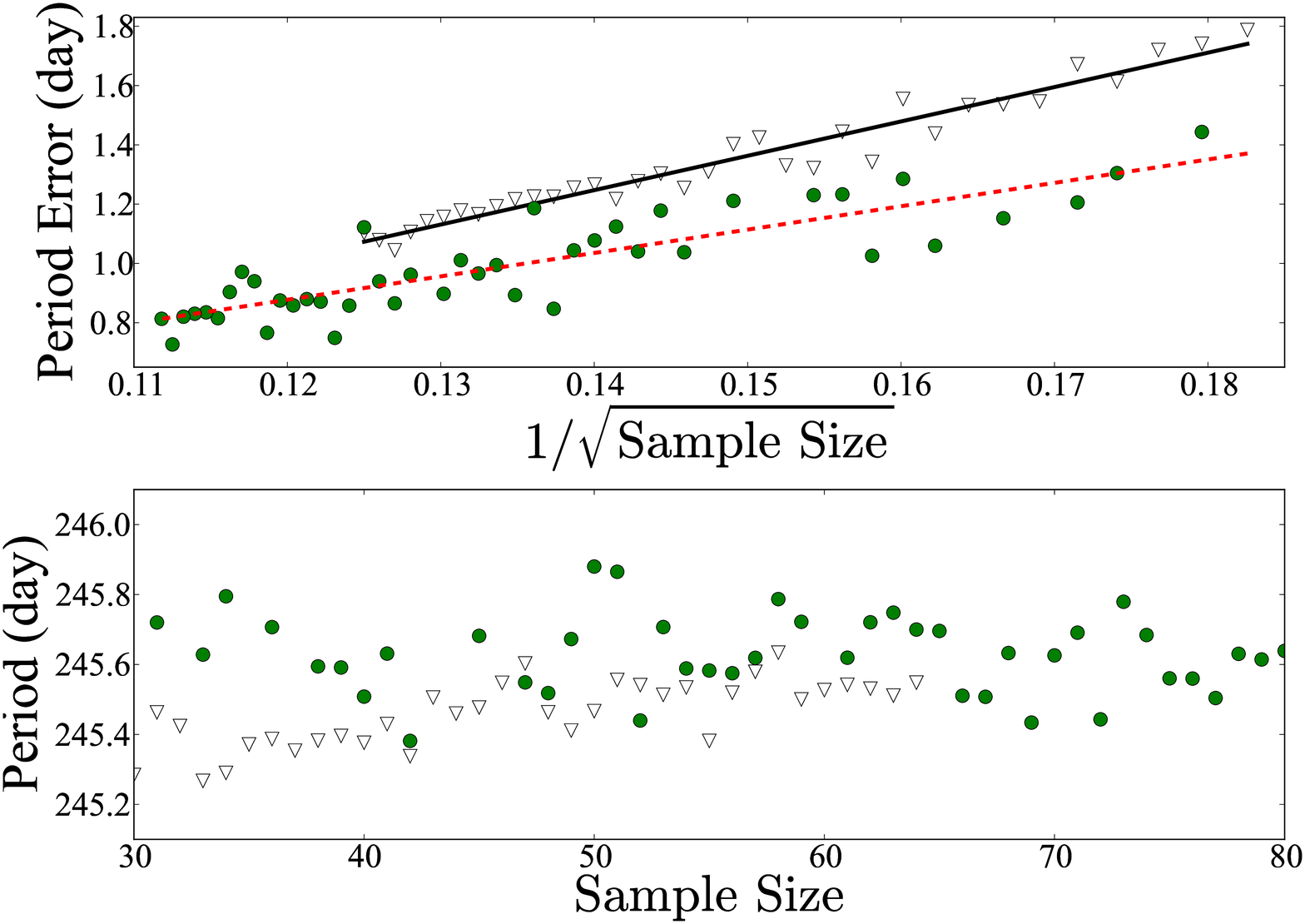}}
\caption{
The top panel is for the measured uncertainties in the periods of the time series of the -89.137\kms\ maser feature as a function of one over square-root of a sample size. For each sample size, the uncertainty in the period was estimated as the standard deviation of the determined periods from the asymmetric cosine fit model to the different combinations of the data points (triangle-down points) and to the randomly uniformly sampled time-stamps which were used together with residuals of the best fitted  asymmetric cosine to the observed time series, to generate the synthetic time series (solid circle points). The solid and dashed line plots are the fitted first order polynomials to the triangle-down and solid circle data points, respectively. The bottom panel is for the mean period in each sample size. The triangle-down points correspond to Monte-Carlo simulation where the time series were randomly selected from 66 population. The solid circles are for the Monte-Carlo simulation where the time-stamps were sampled from a randomly uniform distribution.}
\label{fig:G339.986-0.425_6ghz_error_vs_sample_size}
\end{figure}

\begin{table}
 \begin{minipage}{80mm}
  \caption{
   A summary of the determined periods by fitting our asymmetric cosine model to the time series as the sample size is increased. The periods in column 3, are from the Monte-Carlo simulation where in each $k$ sample size, different combination of time series were selected. The periods in column four are from the randomly uniformly sampling time-stamps to generate a synthetic time series from which our asymmetric cosine model was used to determine the period.
  }
  \label{tab:periods_list_increasesamplesize}
  \begin{center}
  \begin{tabular}{@{}cccc@{}}
  \hline
  Maser feature velocity & Frequency  & \multicolumn{2}{c}{Period}   \\
    (\kms)               & (GHz)      & \multicolumn{2}{c}{(days)}   \\
    \hline   
    -91.068              &  6.7       & 249   $\pm$ 2   & 249   $\pm$ 2   \\
    -89.400              &  6.7       & 246.2 $\pm$ 0.9 & 246   $\pm$ 1   \\     
    -89.137              &  6.7       & 245   $\pm$ 1   & 245.7 $\pm$ 0.9 \\
    -88.610              &  6.7       & 246.6 $\pm$ 0.9 & 246   $\pm$ 1   \\
    -87.907              &  6.7       & 245   $\pm$ 1   & 245   $\pm$ 1   \\
    -89.400              &  12.2      & 246   $\pm$ 1   & 245   $\pm$ 1   \\
    -89.496              &  12.2      & 246   $\pm$ 1   & 246   $\pm$ 1   \\
    -89.160              &  12.2      & 245   $\pm$ 1   & 245   $\pm$ 1   \\
    -88.679              &  12.2      & 243   $\pm$ 2   & 245   $\pm$ 2   \\
    -88.583              &  12.2      & 244   $\pm$ 2   & 245   $\pm$ 1   \\
    \hline
\end{tabular}
\end{center}
\end{minipage}
\end{table}

As already noted, the observed time series give us only one sampling of the true light-curve to which we fitted an asymmetric cosine function to estimate, amongst other things, the period. To find the statistical error on the period from fitting an asymmetric cosine to the light-curve a large number of independent sampling of the same size of the light-curve is necessary. For each sampling the period can be estimated by fitting an asymmetric cosine and by repeating this procedure for a large number of different samplings, it is possible to find the mean and standard deviation of the estimated periods. To by-pass the problem that the observed 6.7- and 12.2-GHz time series each represent only one specific sampling of the respective light-curves, we note that the sampling data allows us to construct a reduced time series, i.e. a time series with fewer sampling points than the total number of sampling points. For example, in the case of the 6.7-GHz masers the total time series consists of 66 time-stamps for which there are $m = 66!/k!(66-k)!$ combinations of time-stamps that can be used to construct a reduced time series consisting of $k < 66$ time-stamps. For each combination an asymmetric cosine can be fitted to the reduced time series to estimate the period. Using all $m$ combinations result in $m$ different periods from which a mean period and standard deviation for a time series with $k < 66$ time-stamps, can be calculated. It is then possible to investigate the behaviour of the error as a function of $k$ for $k \leq 64 = (n-2)$ and extrapolate that result to $k = n = 66$.

The total number of combinations, $m$, can be very large, which, from a computational point of view, makes it unpractical to use all $m$ combinations. We have therefore limited $m$ to 500 and applied the above procedure for $30 \leq k \leq 64$ for the 6.7-GHz masers and $30 \leq k \leq 68$ for the 12.2-GHz masers. Figure \ref{fig:G339.986-0.425_6ghz_error_vs_sample_size} shows how the period (bottom panel) changes as a function of $k$ for the 6.7-GHz maser feature at -89.137\kms. Using the Central Limit Theorem as a guide, we found that the error is proportional to $1/\sqrt{k}$ with the corresponding result shown in the top panel of Figure \ref{fig:G339.986-0.425_6ghz_error_vs_sample_size}. The behaviour for the other maser features was found to be similar and is therefore not shown.  Since the error in the period is proportional to $1/\sqrt{k}$, we can estimate the error in the period for $k = 66$ and $k = 70$ from an extrapolation of a first order polynomial fitted to the relevant data for each maser feature. The mean period and the extrapolated error on the period for all the maser features shown in Figures \ref{fig:G339.986-0.425_67ghz_timeseries} and \ref{fig:G339.986-0.425_12ghz_timeseries} are given in column 3 of Table \ref{tab:periods_list_increasesamplesize}.

In the third case we applied the above procedure to synthetic time series simulated by randomly generating $k-2$ uniformly distributed time-stamps between the start and end MJD of the observed time series. The remaining two time-stamps were the start and end times of the real time series.  Fixing the end point implies that every time series spanned 1015 days. The light-curve was assumed to be given exactly by the asymmetric cosine fitted to the real data. Two sources of scatter around the mean were introduced. The first was to randomly sample from the observed experimental errors. The second was to sample, also randomly, from the residuals between the observed flux densities and the fit from the asymmetric cosine to the real time series.

For each value of $k$, with $30 \leq k \leq 80$, 500 synthetic time series were generated and an asymmetric cosine was fitted to each of them, from which the mean and standard deviation of the period were determined.  The relation between the standard deviation and $1/\sqrt{k}$, and mean period and $k$ as obtained by using fit to the time series of the -89.137\kms\ maser feature are shown in the top and bottom panel of Figure \ref{fig:G339.986-0.425_6ghz_error_vs_sample_size}, respectively. The behaviour observed in Figure \ref{fig:G339.986-0.425_6ghz_error_vs_sample_size} were also noted in the other time series shown in Figures \ref{fig:G339.986-0.425_67ghz_timeseries} and \ref{fig:G339.986-0.425_12ghz_timeseries}.  We can use the fitted first order polynomial to the relation between the standard deviation and $1/\sqrt{k}$ to calculate the standard deviation for sample sizes $k$ = 66 and $k$ = 70 corresponding to the number of time stamps for the real 6.7- and 12.2-GHz time series respectively. A summary of the average period and the standard deviation of the estimated periods is given in the fourth column of Table \ref{tab:periods_list_increasesamplesize}.

It is seen that the uncertainties in the periods obtained with the latter two Monte-Carlo simulations ranges between 0.9 and 2 days. The periods in Table \ref{tab:periods_list_increasesamplesize} are in a very good agreement with each other and with the periods given in Tables \ref{tab:periods} and \ref{tab:asymetric_cosine_periods}. If we assume that the time series shown in Figures \ref{fig:G339.986-0.425_67ghz_timeseries} and \ref{fig:G339.986-0.425_12ghz_timeseries} are statistical independent then the unweighted average period and its uncertainty are 246 $\pm$ 2 and 246 $\pm$ 1 days for the periods in column 3 and 4 of Table \ref{tab:periods_list_increasesamplesize}, respectively.

Since the latter two Monte-Carlo simulations give results that are in a good agreement with each other, we argue further that we can reliably extrapolate the time series to have more than the observed cycles by using the same approach as in our third Monte-Carlo method. This is to investigate how the uncertainty in the period derived from the Lomb-Scargle and epoch-folding methods depends on the number of cycles. The time series of the -89.137\kms maser feature was used for this test. After 40 cycles, the determined periods were 246 $\pm$ 3 and 246 $\pm$ 1 days for the Lomb-Scargle and epoch-folding methods, respectively. The uncertainties in the period are seen to be in good agreement with those estimated above and is in support of an error of about 1 day when fitting an asymmetric cosine to the data. We therefore conclude that the period of the 6.7- and 12.2-GHz masers associated with G339.986-0.425 is 246 $\pm$ 1 days.

%
\subsection{Time Delay }

\begin{figure}
\centering
\resizebox{\hsize}{!}{\includegraphics[clip]{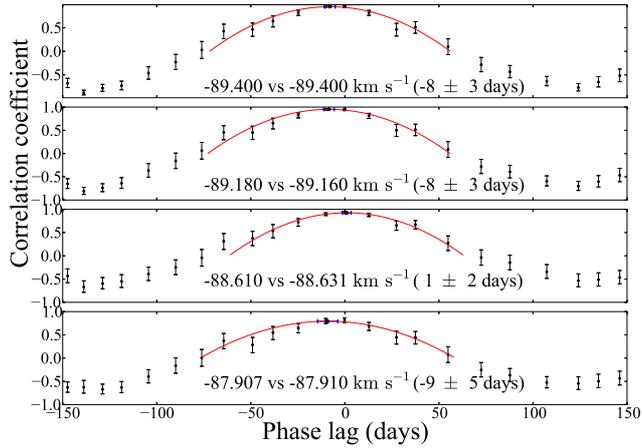}}
\caption{The correlation and time delay results from the ZDCF between the 6.7- and 12.2-GHz masers. The solid lines are for the weighted fitted  second order polynomial on the top half of ZDCF correlation data points. In each panel, the light-curve corresponding to first velocity in the legend was used as a reference light-curve, which was the light-curve of the 6.7-GHz maser feature, and the second velocity was for the 12.2-GHz maser feature. The -89.4\kms\ maser feature was present at both 6.7- and  12.2-GHz. The turning point of the weighted fitted second order polynomial is marked as a star point in each panel.}
\label{fig:G339.986-0.425_6_and_12_timedelay}
\end{figure}

The time series in Figures \ref{fig:G339.986-0.425_67ghz_timeseries} and \ref{fig:G339.986-0.425_12ghz_timeseries} suggest that there are correlations between the 6.7- and  12.2-GHz masers, as well as across the channels of each maser. The correlation and time delay between pairs of time series were tested with the ZDCF. The results for the correlation and time delay between the two masers (6.7- and 12.2-GHz) of a few selected velocities are shown in Figure \ref{fig:G339.986-0.425_6_and_12_timedelay}. The location of a peak in Figure \ref{fig:G339.986-0.425_6_and_12_timedelay}, which was the time delay, was determined from the turning point of a second order polynomial weighted chi-square fit. The errors in the fitted parameters were estimated from the diagonal of the covariant matrix's inverse \citep{press1989}.  The error in the location of the turning point, which is the time delay, was estimated from the errors of the second order polynomial coefficients using propagation of error formula.

\begin{figure}
\centering
\resizebox{\hsize}{!}{\includegraphics[clip]{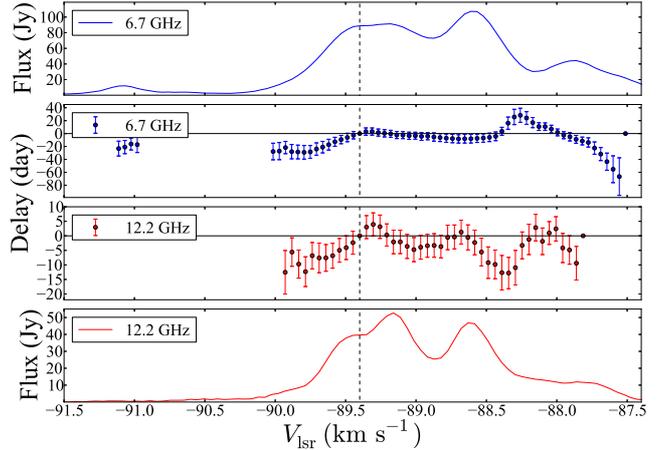}}
\caption{The time delays between some of the selected bright velocities with the -89.400\kms\ in the 6.7- and 12.2-GHz used as a reference time series. The vertical (dashed) lines mark the reference velocity. The horizontal (solid) lines time delays panels mark the zero day point. The average spectra for the 6.7- and  12.2-GHz masers are given in the first and last panels respectively.}
\label{fig:G339.986-0.425_timedelay}
\end{figure}

A summary of the determined time delay as a function of velocity for both 6.7- and 12.2-GHz is shown in Figure \ref{fig:G339.986-0.425_timedelay}. The -89.400\kms\  was present at both 6.7- and 12.2-GHz spectra, and was therefore used as a reference time series for each maser. It is seen in Figure \ref{fig:G339.986-0.425_timedelay} that for both 6.7- and 12.2-GHz masers the delays are negative for the velocities less than -89.400\kms. The 6.7-GHz maser starts with  a delay of about -20 days delay at around 91.0\kms. The delay decreases up to the reference velocity, then starts to increase into positive days to reach the local minimum. After the local minimum, the delay decreases to zero days. Up to this point, the general behaviour for both 6.7- and 12.2-GHz maser time delays is similar. From this point, the 6.7- and 12.2-GHz are different but both show well defined structure which appear to be continuous. One of the noticeable difference between the 6.7- and 12.2-GHz maser delays is around -88.3\kms\ where the 12.2-GHz is at local minimum and the 6.7-GHz is almost at its absolute maximum. For velocities  greater than -88.0\kms, both shows an increase in the delays with 6.7-GHz increasing to about -66 $\pm$ 29 days. The delays for the 6.7- and 12.2-GHz  masers range from -66 to 28, and from -12 to 4 days respectively.

%
\subsection{Maser spot distribution}

\begin{figure}
\centering
\resizebox{\hsize}{!}{\includegraphics[clip]{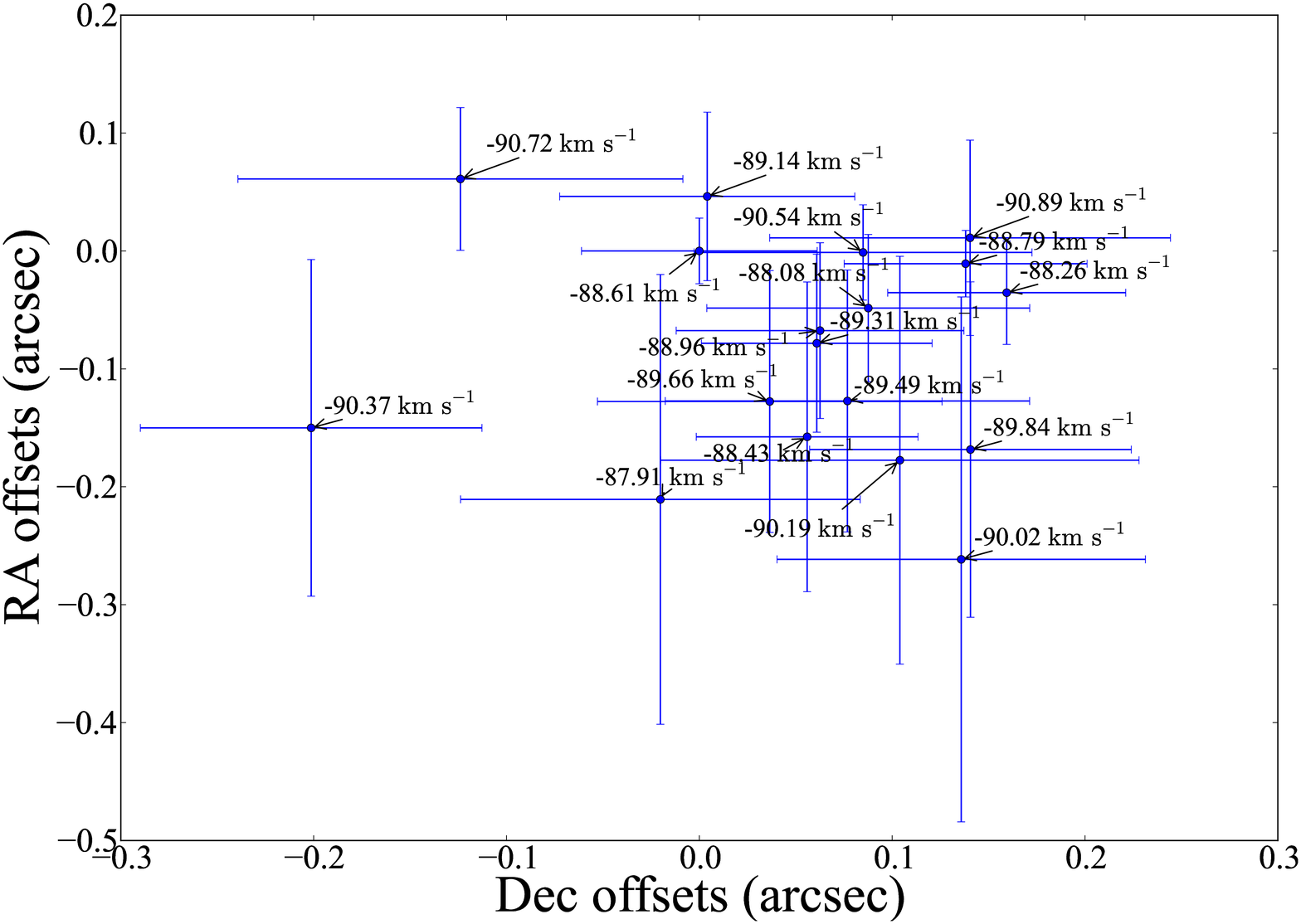}}
\caption{Methanol maser spot map associated with G339.986-0.425 at 6.7-GHz. The reference maser feature was at  -88.61\kms.}
\label{fig:G339.986-0.425_67_spotmap}
\end{figure}

In order to find the maser spot distributions and attempt to link it with time delays shown in Figure \ref{fig:G339.986-0.425_timedelay}, the 6.7-GHz masers associated with the G339.986-0.425 data cube was deconvolved with Common Astronomy Software Applications (CASA). The positions of the maser features in each channel with emission were calculated with AIPS task JMFIT which output the coordinates of the positions and their uncertainties. The declination (DEC) and right ascension (RA) offsets were calculated from the following equations:
\begin{equation}
\begin{aligned}
   &\text{RA}_{offset} = \left (\text{RA} -  \text{RA0}\right)\cos{( \text{DEC0})},\\
  &\text{DEC}_{offset} =  \text{DEC} -  \text{DEC0}, 
  \end{aligned}
  \label{eq:coordinates_offset}
\end{equation}
where DEC0 and RA0 are the position coordinates of the reference channel. The DEC and RA offsets plot for the maser spots distribution is shown in Figure \ref{fig:G339.986-0.425_67_spotmap}. The distributions do not show a simple morphology such as linear, elliptic or circular structures. The uncertainties of the maser spot coordinates suggest that the source is not resolved with ATCA.

\begin{figure}
\centering
\resizebox{\hsize}{!}{\includegraphics[clip]{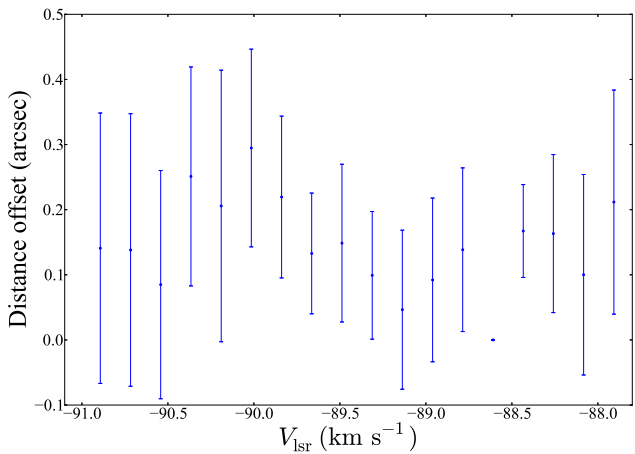}}
\caption{The velocity gradient associated with G339.986-0.425 at 6.7-GHz.}
\label{fig:G339.986-0.425_67_distance_offset}
\end{figure}

As shown in Figure \ref{fig:G339.986-0.425_timedelay} there are time delays across the channels which could suggest a structure which is correlated with velocity. In order to test for possible links between the measured time delays and maser spots relative spatial distribution, the velocity versus distance offset plot was made to test for velocity gradient and the results are shown Figure \ref{fig:G339.986-0.425_67_distance_offset}. The uncertainties in the distance offsets were derived from the propagation of error formula. The offset distances were calculated from the DEC and RA offsets using the distance formula given by equation
\ref{eq:distance_offset}.
\begin{equation}
 \text{Distance offset} = \sqrt{{\text{RA}_{offset}}^2 + {\text{DEC}_{offset}}^2}.
 \label{eq:distance_offset}
\end{equation}
The error in the reference distance offset was undefined as the distance of the reference channel is zero. The uncertainty on the reference channel was therefore set to zero. The positional error in Figure \ref{fig:G339.986-0.425_67_distance_offset} are too large to draw a conclusion about a velocity gradient of the 6.7-GHz masers associated with G339.986-0.425.

%
\section{Discussion}

%
\subsection{Possible origin of the observed periodicity}

As already noted earlier, a number of hypotheses have been made in the past to explain the periodic behaviour of the methanol masers. The question now is whether the periodic behaviour of the masers in G339.986-0.425 can be explained within the framework of any these proposals. The main difficulty in trying to answer this question is that most of these proposals only broadly suggest a particular mechanism but do not predict a specific flare profile or waveform. At present only the colliding-wind binary model of \citet{vanderwalt2011} allows for the explicit calculation of a flare profile. For two other proposals, i.e. that of \citet{parfenov2014} and \citet{inayoshi2013}, is it possible to infer possible flare profiles or light-curves? As for the model of \citet{parfenov2014}, these authors explicitly state that their model is aimed to reproduce the characteristic features of the maser flares as seen in, for example, G9.62+0.20E. The flaring behaviour in G9.62+0.20E is, however, quite different from that of G339.986-0.425 and we therefore will not consider this scenario any further.

The second scenario is the colliding-wind binary model of \citet{vanderwalt2011}. Although this model was originally developed to explain the flaring behaviour of the masers in G9.62+0.20E, \citet{vanderwalt2011} also applied it to the smaller amplitude periodic masers in G188.95+0.89. Using the toy model of \citet{vanderwalt2011} we explored the parameter space to try to produce a light-curve similar to that of G339.986-0.425. Light-curves or flare profiles that very much resembles that of G339.986-0.425 could be found for orbits with an eccentricity of about 0.3. However, lowering the eccentricity necessarily lowers the amplitude of the flaring behaviour and in all these cases it was not possible to reproduce the amplitude of the variations seen in G339.986-0.425. Lowering the eccentricity, reduces ionisation shock front variation range. This is expected to results in a small amplitude variations range. We therefore conclude that also the CWB model cannot explain the periodic maser in G339.986-0.425.

The third scenario is that of \citet{inayoshi2013}, which suggests that the observed flaring is due to a pulsating young high-mass star. As in the case of \citet{parfenov2014}, these authors do not explicitly present a predicted light-curve. However, the underlying mechanism for pulsationally unstable is, according to \citet{inayoshi2013}, the $\kappa$-mechanism which is the same as for many other pulsating stars. It is therefore reasonable to expect the light-curve of the pulsating high-mass star to be similar to that of other pulsating stars also driven by the $\kappa$-mechanism like e.g. Cepheids and the RR Lyrae stars \citep[e.g., ][]{szekely2007,ripepi2015}.  Furthermore, we note that there are OH/IR stars with periods similar to that of G339.986-0.425 and that the light-curves of the associated OH masers can also be fitted quite well with asymmetric cosines \citep{david1996,etoka2000} as is also the case for G339.986-0.425.

Thus, considering the shape of the light-curve of the methanol masers in G339.986-0.425 we conclude that it is compatible with what is expected from an underlying pulsating high-mass star. Adopting the results of \citet{inayoshi2013} the mass of the star can be calculated from their equation 2 and in this case is found to be about 23 solar masses.

%
\subsection{Time delay}

The question may arise whether the measured delays can be explained under the consideration of the YSO's pulsational instability as the origin of the periodicity. The masing cloudlets separations around high-mass YSO are very small compared to the distance between the maser cloudlets and the observer. This implies that the emissions from masing cloudlets in the region will travel approximately equal distances to the observer. The measured time delay in the maser time series could be due to their relative spatial positions to the pulsating YSO. This ignores the effects of radiation reprocessing. If the masing cloudlets and YSO are projected onto a 2-d plane, then the time delays give the least separation distances of the masing cloudlets. 

If the disturbances travel with the speed of light in a vacuum, an 8 $\pm$ 3 day time delay between the 6.7- and 12.2-GHz at -89.400\kms\ implies that these masers are (1.4 $\pm$ 0.5)\e{3} AU apart, which is larger than the diameter of the Solar system. Since the velocities are the same, the kinematics around these masers are more likely to be similar because the masers are expected to be close to each other in the molecular cloud. For a sound speed in a molecular cloud of $\sim$ 0.5\kms, derived using equation 15.22 in \citet{kwok2007} with the temperature assumed to be 100 K, and the disturbances travel at sound speed, then an 8 $\pm$ 3 day time delay implies that the -89.400\kms\ maser features at 6.7- and 12.2-GHz are $\sim$ (2.3 $\pm$ 0.7)\e{-3} AU apart. The separation is quite small. However, from the maser spot map (Figure \ref{fig:G339.986-0.425_67_spotmap}) we approximate the total extent of the maser emission angular separation to be 0.4 arcseconds. The Galactic longitude of G339.986-0.425 with a maser peak velocity of -89.0\kms\ can be used to derive the near kinematic distance to the source using the rotation curve of \citet{wouterloot1989}, and it is $\sim$ 5.6 kpc. Using the derived near kinematical distance to G339.986-0.425, $\sim$ 5.6 kpc, and the total extent of the maser emission angular separation of $\sim$ 0.4 arcseconds, the diameter of the total extent of the maser emission was calculated to be $\sim$ 0.01 pc or $\sim$ 12 light days. The diameter of the 6.7-GHz maser emission in G339.986-0.425 was found to be in good agreement with the typical methanol maser region linear size of 0.003 pc for the near kinematics distances of 6 kpc derived by \citet{caswell1997}. If the disturbances travel with  a sound speed of $\sim$ 0.5\kms, it will take about 2\e{4} years to travel the distance of 0.01 pc. The maximum time delay measured in Figure \ref{fig:G339.986-0.425_timedelay} is 66 $\pm$ 28 days. In the case where the disturbances travel with the speed of light in a vacuum, the separation distance for the maser features would be (1.1 $\pm$ 0.5)\e{4} AU or 0.06 $\pm$ 0.02 pc which is larger than the approximated total masing cloudlets diameter. If the disturbances travel at sound speed then the separation between these maser features would be (1.9 $\pm$ 0.8)\e{-2} AU, which is much smaller than 0.01 pc. From this analysis, it is clear that if the disturbances travel at sound speed, then the total extent of the masing region is in many orders of magnitude too small for a typical methanol masing region of about 0.003 pc \citep{caswell1997}. Given that we see the masers in projection that strongly suggests that it is either the seed photons or something directly related to radiation that is causing the periodic changes. 

The measured delayed time delays in G339.986-0.425 are not easy to explain. For a better understanding of such large delays and the time delays structure seen in Figure \ref{fig:G339.986-0.425_timedelay}, interferometric data with a similar spectral resolution as the one obtained with the 26m HartRAO radio telescope and high spatial resolution of the source is vital. The spectral resolution of the ATCA data was 1.25 greater than the 26m HartRAO radio telescope data.

%
\section{Summary}

A search for periodic methanol masers associated with high-mass YSOs was conducted with the 26m HartRAO telescope  and ten methanol masers from the  6.7-GHz MMB survey catalogues III and IV were selected as the candidates for the search. The methanol maser associated with G339.986-0.425 at 6.7- and 12.2-GHz was the only source from ten source sample to show periodic variations with a 246 $\pm$ 1 day period. 

The time series of the two brightest class II methanol masers associated with G339.986-0.425 show strong correlations and time delays between the maser species and across the velocity channels. The calculated time delays show remarkable structure when plotted against velocities. The meaning of the structure is not clear as the interferometric maser distribution did not help, mainly due to, insufficient spatial and spectral resolution. Another remarkable behaviour was the time delay of 8 days between the -89.400\kms\ feature for both 6.7- and 12.2-GHz  masers. The velocity was the same for both masers which suggests that the kinematics could be similar if not the same.   

The catalogue for periodic masers has been increased to sixteen sources and will certainly improve the statistical analysis  of these sources. The G339.986-0.425 light-curves had lead to the proposal that high-mass YSO pulsational instability could be the origin of the periodicity. However, the origin of the periodic variations in maser is still to be confirmed. Numerical modelling could possibly help in narrowing the possibilities or introduce possible new model in the origin of the observed periodicity in methanol masers, e.g., testing the maser-response to the wide variety of time-dependent dust temperature light-curves.

\section*{Acknowledgments}
Jabulani Maswanganye would like to express his gratitude for the bursaries from the Hartebeesthoek Radio Astronomy Observatory, the North-West University doctoral scholarships and the South African SKA Project via the NRF.

\bsp

\label{lastpage}

\end{document}